\documentclass[11pt]{article}
\usepackage{geometry}                
\usepackage[parfill]{parskip}    
\usepackage{graphicx}
\usepackage{amssymb}
\usepackage{epstopdf}
\fussy
 \title{Using Jet Asymmetries to Access $\Delta$G at STAR}
 \author{Renee Fatemi for the STAR Collaboration\\Massachusetts Institute of Technology}
 \date{}

\begin{document}
 \maketitle
 \begin{abstract}
We present the first inclusive jet longitudinal double spin asymmetry $A_{LL}$ and differential cross-section measurement from STAR. This analysis reflects 1 $pb^{-1}$  of polarized $\vec{p}\vec{p}$ data taken at $\sqrt{s}=200$ GeV with beam polarizations of 30-45$\%$ in years 2003 and 2004.  Agreement between the measured differential cross-section and next-to-leading order  (NLO) calculations validate the interpretation of the asymmetry results within the framework of perturbative Quantum Chromodynamics (pQCD).  $A_{LL}$ is compared to NLO pQCD  calculations,  which incorporate several possible gluon helicity distributions, and is shown to disfavor the maximum gluon scenario. 
 \end{abstract}
 
One ongoing question of QCD concerns how the spin of the partons, which compose the proton, contribute to the total proton spin. Conservation of angular momentum requires:
\begin{equation}
J_{proton}=\frac{1}{2} = {\Delta}S_q + {\Delta}G + L_s + L_G
\end{equation}
where ${\Delta}S_q$ is the spin of the quarks, ${\Delta}$G the spin of the gluons and $L_q$ and $L_G$  the angular momentum of the respective partons.  Inclusive fixed target polarized lepton-nucleon scattering experiments in the deep inelastic scattering (DIS) region have mapped out ${\Delta}S_q$ as a function of  Bjorken x and $Q^2$\cite{DIS}, constraining the contribution to 10-30$\%$ of the total spin of the proton.  Lepton scattering provides access to the gluon spin only at next-to-leading order (NLO), requiring fits to the world dataset of the first spin structure function of the proton $g_1^p(x,Q^2)$ in order to extract $\Delta G$. One such analysis results  in  $\Delta G(Q^2=1 GeV) = 0.99 ^{+ 1.17}_{-0.31}$\cite{SMC}, where the limited precision is directly related to the available phase space of the fixed target experiments.  Experimentally there are two ways to reduce this error, the first requires a polarized lepton-proton collider which would allow access to a larger x and $Q^2$ region. The second  possibility, currently being pursued at the Relativistic Heavy Ion collider (RHIC),  is to probe the gluon directly using strong interactions. RHIC is the worlds first polarized proton-proton collider, providing collisions at energies of ${\sqrt{s}=50-500}$ GeV with average luminosities of 10${^{30}}$-10${^{35}}$ and polarizations up to 60${\%}$. The STAR (Solenoidal Tracker at STAR) Spin Collaboration at RHIC has initiated a multi-year program devoted to measurements which facilitate the extraction of the polarized gluon distribution in the proton. The first stage of this program includes measurements of inclusive jet, charged and neutral hadron longitudinal double spin asymmetries ($A_{LL}$) and differential cross-sections with the ultimate goal being to map the x dependence of the gluon helicity distribution using the $\gamma+jet$ channel.  These proceedings present an important step towards this  goal, the first inclusive jet $A_{LL}$ and differential cross-section measurements from STAR. 

STAR \cite{STAR} was designed to be a large acceptance detector, optimized for mid-rapidity jet reconstruction. The detector systems relevant for  jet reconstruction are the Time Projection Chamber  (TPC) and the Barrel (BEMC) and Endcap (EEMC) Electromagnetic Calorimeters.  The TPC provides momentum information on charged particles scattered in the pseudorapidity region $|\eta|<1.4$.   The BEMC and EEMC measure the neutral energy  deposited per event in the range spanning $-1<\eta<2$. The calorimetry was fully installed and incorporated into the trigger in time for the 2006 run, therefore the results reported here use only the calorimetery in region of $0<\eta<1$. The minimum bias trigger (MINB) was defined by a coincidence signal between East and West Beam Beam Counters (BBC) \cite{BBC}, which are  segmented scintillator detectors located on either side of the interaction region at $3.3 <|\eta|<5.0$. Enhancement of jet reconstruction at  high transverse momentum (pT)  was achieved by using a High Tower (HT) trigger in coincidence with the MINB trigger. The HT trigger requires the deposition of  $E_T>2.2-3.2$ GeV in a single tower ($\Delta\eta\times\Delta\phi = 0.05\times0.05)$. Finally the BBC also serves as the STAR luminosity monitor, providing the spin dependent luminosity normalization for the asymmetry analysis. 

The STAR jetfinder is based on the collinear and infrared safe Midpoint Cone Algorithm\cite{ALGO}. This algorithm clusters TPC tracks and BEMC towers into jets with a radius of R=$\sqrt{\Delta\eta^2+\Delta\phi^2}$ =0.4 and minimum pT of 5 GeV. A seed of 0.5 GeV is required to begin clustering and jets with more than 50$\%$ of their energy overlapping are merged. The standard set of jet  cuts for  the  2003(2004)  analysis include a vertex requirement  of  $|Z_{ver}|<75 (60)$ to ensure uniform TPC efficiency, a BEMC fiducial cut which limits the jet axis to point at least 0.2 units in $\eta$ from the edge of the detector in order to minimize edge effects and the requirement that the jet neutral energy fraction  be $R_{ET}< 0.8(0.9)$  to reduce contamination from beam background. The cross-section analysis also applies an offline software trigger requiring the trigger tower contain $E_T > 3.5$ GeV.


\begin{figure}[t]
\unitlength1.cm
\center
\includegraphics[height=11cm,width=9.0cm]{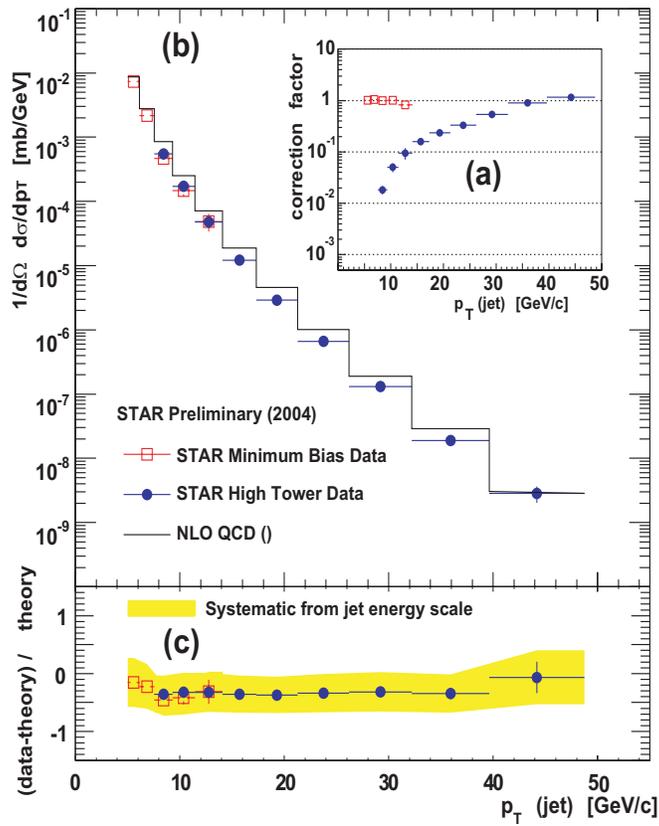}
\caption{\label{XSECfig}a) The factors $c(p_T)$, derived from simulations, which are used to correct the raw jet yield for trigger, detector and jet reconstruction inefficiencies. b) The 2004 inclusive jet differential cross-section compared to NLO pQCD calcuations \cite {XSECnlo}. C) The yellow band represents the systematic error due to to the uncertainty in the jet energy scale.}
\end{figure}

The extraction of the gluon helicity distribution function from spin asymmetries requires that pQCD is applicable to the jets that are reconstructed at RHIC pT and x ranges. The measurement and comparison of the inclusive jet cross-section to NLO predictions provides this crucial test. The differential cross-section was determined according to:  
\vspace{-0.1cm}
\begin{equation}
\frac{d^2\sigma}{d{\Omega}dp_T}=\frac{N_{jets}}{\Delta\eta\Delta\phi\Delta{p_T}}\cdot\frac{1}{\int{Ldt}}\cdot\frac{1}{c(p_T)}
\end{equation}
\vspace{-0.1cm}where $c(p_T)$ is a correction factor which accounts for detector, trigger and jet reconstruction inefficiencies. These factors are determined using the STAR simulation package, consisting of  PYTHIA 6.205 set to the "CDF Tune A" parameters  and a GEANT detector simulation.  The inset figure  in the upper right hand corner of Fig.\ref{XSECfig} demonstrates that, as expected,  the MINB c$(p_T) $are flat in pT and nearly unity. However, due to trigger inefficency, the high-tower correction factors vary by 2 orders of magnitude, reaching corrections as small as 5 x10$^-3$ at pT= 5  GeV. The lowest pT bins are removed from the high tower sample due to the unacceptably large systematic errors resulting from dividing by correction factors  less than 0.01. This strong pT dependence has been reduced in later runs via the implementation of a less biased Jet Patch trigger. 

The inclusive jet differential cross-section,  representing a sampled 0.2 $pb^{-1} $ of data  from 2004 only,  is compared to NLO predictions\cite{XSECnlo} in Fig. \ref{XSECfig}. The minbias  and high tower samples show excellent agreement in the region of overlap, and in general there is good agreement with NLO over 7 orders of  magnitude.  While the STAR cross-section is systematically lower than NLO by ~$30\%$, this is within the dominant systematic due to the calibration of the jet energy scale.  The relative calibration of the BEMC was completed $\it{in}$ $\it{situ}$ using the response of the towers to charged tracks in d+Au (Au+Au) collisions in 2003 (2004). The absolute energy scale was set by matching the BEMC tower response to electrons, selected via TPC dE/dx cuts,  which traverse the center of the tower.  A conservative estimate of $10\%$ uncertainty in calorimeter calibration results in a $50\%$ uncertainty in the yield.  STAR plans to use pion reconstruction and $\gamma+jet$ events to reduce the calibration uncertainty in the future. 

\begin{figure}[t]
\unitlength1.cm
\center
\includegraphics[height=7.0cm,width=10.cm]{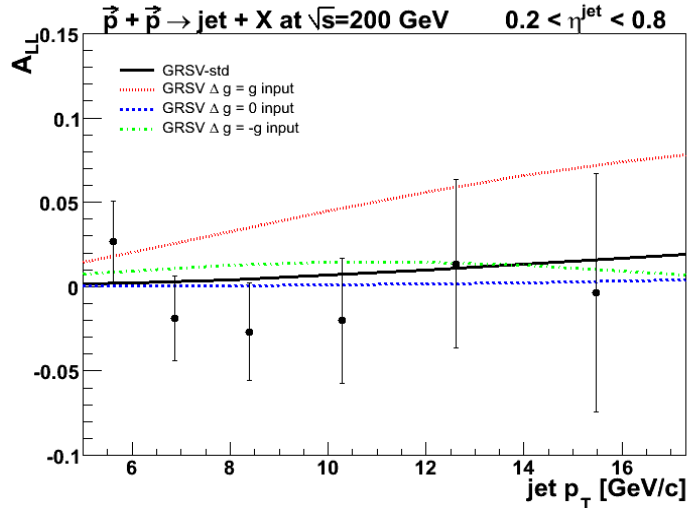}
\caption{\label{ALLfig}The longitudinal double spin inclusive jet asymmetry $A_{LL}$ for years 2003+2004 compared to NLO pQCD calculations\cite{XSECnlo}.  The curves incorporate four possible gluon helicity distributions allowed from fits to experimental data : $\Delta{g}=g$ (red), $\Delta{g}=-g$ (green), $\Delta{g}=0$ (blue), $\Delta{g}=standard$ (black).}
\end{figure}

In 2003 and 2004 both RHIC polarized proton beams were composed of 55 bunches, each assigned a distinct helicity state. The helicity pattern of one beam alternated with each bunch and the other every two bunches, resulting in four possible helicity combinations at the interaction region (++,+-,-+,- -).  Incorporating this spin information on an event by event basis, $A_{LL}$ is calculated according to:
\vspace{-0.1cm}
\begin{equation}
A_{LL}=\frac{1}{P_1P_2}\frac{N^{S}  - N^{O}}{N^{S}+N^{O}}
\end{equation}
\vspace{-0.1cm}where $N^{S(O)}$ is the luminosity normalized jet count for collisions of the same (oppositie) proton beam helicity states. The factor  ${P_1P_2}$ is  the product of the two beam polarizations, varying from 0.1-0.25, which were measured on a fill-by-fill basis using Carbon nuclear-interference fast carbon-proton polarimeters \cite{CNI}. This factor carries a scale uncertainty of $25\%$ which is not included in the statistical errors on $A_{LL}$.  Figure \ref{ALLfig}  shows the combined 2003+2004 inclusive jet double spin asymmetry results in comparison with NLO predictions for $A_{LL}$ \cite{XSECnlo} which incorporate several possible scenarios for $\Delta{G}$\cite{GRSV}.  This data sample is statistically limited and cannot discern between the standard, zero and negative gluon helicity scenarios.  It does however disfavor the maximal case, where the gluon helicity distribution is equal to the unpolarized gluon distribution. This result is consistent with previous evaluations of $\Delta{G}$ from DIS experiments. The point to point systematic error for $A_{LL}$ is 0.01, less than the statistical errors at the lowest pT bin, and is not included in Figure \ref{ALLfig}. The dominant systematic errors include contributions from the relative luminosity (0.009), residual non-longitudinal spin asymmetries (0.01), beam background (0.003) and trigger bias (0.007). Additional checks for statistically significant  non-zero parity violating asymmetries and randomized bunch to bunch helicity distributions give no indication of beam related systematics.

We  have measured the inclusive jet double spin asymmetry $A_{LL}$ and differential cross-section at $\sqrt{s}=200$ GeV. The cross-section, although limited  by a significant systematic error on the jet energy scale, agrees well over 7 orders of magnitude with NLO pQCD calculations and supports the application of pQCD to the asymmetry results. Future improvements in calorimeter calibration may allow a reduction of this systematic error, and thereby allow additional contraints on the high x gluon unpolarized parton distribution functions. Comparison of $A_{LL}$ with NLO pQCD calculations show that the measured asymmetry disfavors the maximal gluon helicity distribution scenario.


\begin{thebibliography}{0}
\vspace{-0.3cm}
\bibitem{DIS} B. Filippone and X.D. Ji,The Spin Structure of the Nucleon, Adv. Nucl. Phys. 26:1, 2001.
\vspace{-0.3cm}
\bibitem{SMC} SMC Collaboration Phys.Rev. D58, 112002, 1998.
\vspace{-0.3cm}
\bibitem{STAR} Special Issue: RHIC and Its Detectors, Nucl. Instrum. Meth. A499, 2003.
\vspace{-0.3cm}
\bibitem{BBC} J.Kiryluk {\it et al.}, hep-ex/0501072, published in Spin 2004 Conference Proceedings, Trieste, Italy.
\vspace{-0.3cm}
\bibitem{ALGO} G.C. Blazey {\it et al.}  arXiv:hep-ex/0005012
\vspace{-0.3cm}
\bibitem{XSECnlo} B. Jager, M. Stratmann, W. Vogelsang, Phys. Rev. D70 034010, 2004.
\vspace{-0.3cm}
\bibitem{CNI} O. Jinnouchi  {\it et al.}, RHIC/CAD Accelerator Physics Note 171, 2004.
\vspace{-0.3cm}
\bibitem{GRSV}M. Gl\"{u}ck, E. Reya, M. Stratmann and W. Voglesang, Phys. Rev. D69, 094005, 2001.
\end{thebibliography}
 \end{document}